\documentclass[12pt]{article}
\begin{document}
\pagestyle{plain}
\title{\huge \bf  Energy shift of H-atom electrons due to Gibbons-Hawking thermal bath}
%\large
\author{Miroslav Pardy\\[7mm]
%Institute of Plasma Physics ASCR\\
%Prague Asterix Laser System, PALS\\
%Za Slovankou 3, 182 21 Prague 8, Czech Republic\\
%and\\
Department of Physical Electronics \\
and\\
Laboratory of Plasma Physics\\[5mm]
Masaryk University \\
Kotl\'{a}\v{r}sk\'{a} 2, 611 37 Brno, Czech Republic\\
e-mail:pamir@physics.muni.cz}
\date{\today}
\maketitle

\vspace{20mm}
%\large
\begin{abstract}
The electromagnetic shift of energy levels  of H-atom electrons is determined by calculating an electron coupling to the 
Gibbons-Hawking electromagnetic field thermal bath. Energy shift of electrons in H-atom is determined  in the framework of non-relativistic quantum mechanics. 
\end{abstract}

\vspace{10mm}

%\large

\hspace{3ex}
\baselineskip 17pt
 
The Gibbons-Hawking effect is the statement that a
temperature can be associated to each solution of the
Einstein field equations that contains a causal horizon. It
is named after Gary Gibbons and Stephen William Hawking.

Schwarzschild spacetime contains an event
horizon and so can be associated with temperature. In the
case of Schwarzschild spacetime this is the temperature
$T$ of a black hole of mass $M$, satisfying $T/M$ .

De Sitter space which contains an
event horizon has the temperature $T$ proportional
to the Hubble parameter $H$.
We consider here the influence of the heat bath of the Gibbons-Hawking photons on the energy shift of H-atom electrons.  

The considered problem is not in the scientific isolation, because some analogical problems are solved in the scientific respected journals. At present time it is a general conviction that there is an important analogy between black hole and the hydrogen atom. The similarity between black hole and the hydrogen atom was considered for instance by Corda (2015a), who discussed the precise model of Hawking radiation from the tunneling mechanism. In this article an elegant expression of the probability of emission is given in terms of the black hole quantum levels. So, the system composed of Hawking radiation and black hole quasi-normal modes introduced by Corda (2015b)  is somewhat similar to the semiclassical Bohr model of the structure of a hydrogen atom. 

 The time dependent Schr\"odinger
equation was derived for the system composed by Hawking radiation and black hole quasi-normal modes (Corda, 2015c). In this model, the physical state and the correspondent wave function are written in terms of an unitary evolution matrix instead of a density matrix. Thus,
the final state is a pure quantum state instead of a mixed one and it means that there is no information loss. Black hole  can be well defined as the 
quantum mechanical systems, having ordered, discrete quantum spectra,
which respect 't Hooft's assumption that Schr\"odinger equations can be used universally for all dynamics in the universe. 
 
Thermal photons by Gibbons and Hawking form so called blackbody, which has the distribution law of photons derived in 1900 by Planck (1900, 1901), (Sch\"opf, 1978). 
The derivation was based on the investigation of the statistics of the system of oscillators inside of the blackbody. Later Einstein (1917)  derived the Planck formula from the Bohr model of atom where electrons have the discrete energies and  
the energy of the emitted photons are given by the Bohr formula 
$\hbar\omega = E_{i} - E_{f}$,  $E_{i},  E_{f}$ are the initial and final energies of electrons.  
 
Now, let us calculate the modified Coulomb potential due to blackbody. 
The starting point of the determination of the energy shift in the H-atom is 
the potential $V_{0}({\bf x})$, which is generated by nucleus of the H-atom.
The potential at point $V_{0}({\bf x} + \delta{\bf x})$, evidently is
(Akhiezer, et al., 1953; Welton, 1948):

$$V_{0}({\bf x} + \delta{\bf x}) = \left\{1 + \delta{\bf x}\nabla +
\frac{1}{2}(\delta{\bf x}\nabla)^{2} + ...\right\}V_{0}({\bf x}). \eqno(1)$$

If we average the last equation in space, we can eliminate so called the effective potential in the form

$$V({\bf x})  = \left\{1 + \frac{1}{6}(\delta{\bf x})_{T}^{2}\Delta + ...\right\}
V_{0}({\bf x}), \eqno(2)$$
where $(\delta{\bf x})_{T}^{2}$ is the average value of te square coordinate shift caused by the thermal photon fluctuations. The potential shift follows from eq. (2):

$$\delta V({\bf x})  = \frac{1}{6}(\delta{\bf x})_{T}^{2}\Delta V_{0}({\bf x}).
 \eqno(3)$$

The corresponding shift of the energy levels is given by the standard quantum mechanical formula 
(Akhiezer, et al., 1953)

$$\delta E_{n}  = \frac{1}{6}(\delta{\bf x})_{T}^{2}
(\psi_{n}\Delta V_{0}\psi_{n}). \eqno(4)$$

In case of the Coulomb potential, which is the case of the H-atom,
we have

$$V_{0} = - \frac{e^{2}}{4\pi|{\bf x}|}.\eqno(5)$$

Then for the H-atom we can write 

$$\delta E_{n}  = \frac{2\pi}{3}(\delta{\bf x})_{T}^{2}\frac{e^{2}}{4\pi}
|\psi_{n}(0)|^{2},\eqno(6)$$
where we used the following equation for the Coulomb potential

$$\Delta\frac{1}{|{\bf x}|} = -4\pi\delta({\bf x}).\eqno(7)$$

Motion of electron in electric field 
is evidently described by elementary equation
 
$$ \delta\ddot{\bf x} = \frac{e}{m} {\bf E}_{T},\eqno(8)$$ 
which can be transformed by the Fourier transformation into the following 
equation
 
$$ |\delta{\bf x}_{T\omega}|^{2} = \frac{1}{2}
\left(\frac{e^{2}}{m^{2}\omega^{4}}\right) {\bf E}_{T\omega }^{2},\eqno(9)$$ 
where the index $\omega$ concerns the Fourier component of above functions.

On the basis of the Bethe idea of the influence of vacuum fluctuations on the energy shift of electron (Bethe, 1947), the following elementary relations was used by Welton (1948), Akhiezer et al. (1953) and  Berestetzkii et al. (1999):

$$\frac{1}{2}{\bf E}_{\omega}^{2} = \frac{\hbar\omega}{2}\eqno(10)$$ 
and in case of the thermal bath of the blackbody, the last equation is of 
the following form (Isihara, 1971):

$$  {\bf E}_{T\omega}^{2} 
= \varrho(\omega) = \left(\frac{\hbar\omega^3}{\pi^2 c^3}\right)\frac{1}
{e^{\frac{\hbar\omega}{kT}} - 1},\eqno(11)$$ 
because the Planck law in (11) was  written as 

$$\varrho(\omega) = G(\omega)<E_{\omega}> =
\left(\frac{\omega^2}{\pi^2 c^3}\right)\frac{\hbar\omega}
{e^{\frac{\hbar\omega}{kT}} - 1},\eqno(12)$$
where the term 
$$<E_{\omega}>  = \frac{\hbar\omega} {e^{\frac{\hbar\omega}{kT}} - 1}
\eqno(13)  $$  
is the average energy of photons in the blackbody and 

$$G(\omega) = \frac{\omega^2}{\pi^2 c^3}\eqno(14)$$
is the number of electromagnetic modes in the interval $\omega, \omega
+ d\omega$. 

Then, 

$$(\delta{\bf x}_{T\omega})^{2} =  \frac{1}{2}
\left(\frac{e^{2}}{m^{2}\omega^{4}}\right)
\left(\frac{\hbar\omega^3}{\pi^2 c^3}\right)
\frac{1}{e^{\frac{\hbar\omega}{kT}} - 1},\eqno(15)$$
where  $(\delta{\bf x}_{T\omega})^{2}$ involves the number of frequencies 
in the interval $(\omega, \omega + d\omega)$. 

So, after some integration, we get 

$$(\delta{\bf x})_{T}^{2} = \int_{\omega_{1}}^{\omega_{2}} \frac{1}{2}
\left(\frac{e^{2}}{m^{2}\omega^{4}}\right) 
\left(\frac{\hbar\omega^3}{\pi^2 c^3}\right)
\frac{d\omega}
{e^{\frac{\hbar\omega}{kT}} - 1} = \frac{1}{2}\left(\frac{e^{2}}{m^{2}}\right) 
\left(\frac{\hbar}{\pi^2 c^3}\right)
 F({\omega_{2}}  - {\omega_{1}}),\eqno(16)$$
where $F(\omega)$ is the primitive function of the omega-integral

$$J = \frac{1}{\omega}
\frac{1} {e^{\frac{\hbar\omega}{kT}} - 1},\eqno(17)$$
which cannot be calculated by the elementary integral methods and it is not involved in the tables of integrals.

 Frequencies ${\omega_{1}}$ and ${\omega_{2}}$ will be  determined with regard to the existence of the fluctuation field of thermal photons. It was determined in case of the Lamb shift (Bethe, 1947 ; Welton, 1947) by means  of the physical analysis of the interaction of the Coulombic atom with the surrounding fluctuation field. We suppose here that the Bethe and Welton arguments are valid and so we take the frequencies in the Bethe-Welton form. In other words, electron cannot respond to the fluctuating field if the frequency which is much less than the atom binding energy given by  the Rydberg constant (Rohlf, 1994) $E_{Rydberg}  = \alpha^{2}mc^{2}/2$. So, the lower frequency limit is 

$${\omega_{1}} = E_{Rydberg}/\hbar  =  \frac{\alpha^{2}mc^{2}}{2\hbar},\eqno(18)$$
where $\alpha \approx 1/137$ is so called the fine structure constant.

The specific form of the second frequency follows from the elementary argument, that 
we expect the effective cutoff, since we must neglect the relativistic effect in our non-relativistic theory. So, we write

$${\omega_{2}} =  \frac{m c^{2}}{\hbar}.\eqno(19)$$
 
If we take the thermal function of the form of the geometric series

$$ \frac{1} {e^{\frac{\hbar\omega}{kT}} - 1} =  q(1 +q^{2} + q^{3} +  .....); 
\quad q =  e^{-\frac{\hbar\omega}{kT}},\eqno(20)$$

 $$\int_{\omega_{1}}^{\omega_{2}}q(1 +q^{2} + q^{3} +  .....) 
 \frac{1}{\omega} d\omega = \ln|\omega| + \sum_{k=1}^{\infty} 
\frac{(-\frac{\hbar\omega}{kT})^k}{k!k} + ....; 
\quad q =  e^{-\frac{\hbar\omega}{kT}}\eqno(21)$$ 
and the first thermal contribution is 

$$ Thermal\; contribution =  \ln\frac{\omega_{2}}{\omega_{1}} - 
\frac{\hbar}{kT}({\omega_{2}} - {\omega_{1}}),\eqno(22)$$ 

Then, with eq. (6)

$$\delta E_{n}  \approx \frac{2\pi}{3}\left(\frac{e^{2}}{m^{2}}\right) 
\left(\frac{\hbar}{\pi^{2}c^{3}}\right)
\left(\ln\frac{\omega_{2}}{\omega_{1}}
-\frac{\hbar}{kT}({\omega_{2}} - {\omega_{1}})\right)
|\psi_{n}(0)|^{2}, \eqno(23)$$
where (Sokolov et al., 1962)

$$ |\psi_{n}(0)|^{2} = \frac{1}{\pi n^{2}a_{0}^{2}}\eqno(24)$$ 
with 

$$ a_{0} = \frac{\hbar^{2}}{m e^{2}}.\eqno(25)$$ 

Let us only remark that the numerical form of eq. (23) has deep experimental astrophysical meaning.

In article by author (Pardy, 1994), which is the  continuation of author articles on the finite-temperature \v Cerenkov radiation and gravitational \v Cerenkov radiation (Pardy, 1989a; ibid., 1989b), the temperature Green function in the framework
of the Schwinger source theory was derived in order to determine the 
Coulomb and Yukawa potentials at finite-temperature
using the Green functions of a photon with and
without radiative corrections, and then by considering
the processes expressed by the Feynman diagrams.

The determination of potential at finite temperature is one of
the problems  which form the basic ingredients of the
quantum field theory (QFT) at finite temperature. This theory 
was formulated some years ago by Dolan and Jackiw (1974), Weinberg (1974) and
Bernard (1974) and some of the first applications of this theory were
the calculations of the temperature behavior of the effective
potential in the Higgs sector of the standard model.

Information on the systematic examination of the finite temperature
effects in quantum electrodynamics (QED) at one-loop order was given
by Donoghue, Holstein and Robinett (1985).  Partovi (1994) discussed the QED
corrections to Planck's radiation law and photon thermodynamics,

A similar discussion of QED was published by Johansson, Peressutti
and Skagerstam (1986) and Cox et al. (1984).

Serge Haroche (2012) and his research group in the Paris microwave
laboratory used a small cavity for the long life-time of photon
quantum experiments performed with the Rydberg
atoms. We considered here the  thermal gas corresponding to the Gibbons-Hawking theory of space-time
(at temperature T) as the preamble for new experiments for the
determination of the energy shift of H-atom electrons interacting with the Gibbons-Hawking on thermal gas. 
It is not excluded, that the observations 
performed by the well educated astro-experts will be the Nobelian ones.

\vspace{10mm}

\noindent
{\bf References}
\vspace{5mm}

\noindent
Berestetzkii, V. B., Lifshitz, E. M. and Pitaevskii, L. P.  Quantum electrodynamics, (Butterworth-Heinemann, Oxford, 1999).\\[2mm]
Bethe, H. A. (1947). The electromagnetic shift of energy levels, Phys. Rev.  {\bf 72}, 339.\\[2mm]
Bernard. C. W. (1974). Feynman rules for gauge theories at finite temperature,
 Phys. Rev. D {\bf 9}, 3312.\\[2mm]
Corda, Ch. (2015a). Precise model of Hawking radiation from the tunneling mechanism, 
Class. and Quantum Gravity {\bf 32}, 195007. \\[2mm]
Corda, Ch. (2015b). Quasi-normal modes: the "electrons" of Blak holes as "gravitational 
atoms"? Implications for the black hole information puzzle. 
Advances in High Energy Physics, 867601. \\[2mm]
Corda, Ch. (2015c). Time dependent Schr{\"o}dinger equation for black hole evaporation: no information loss, Annals of Physics {\bf 353}, 71. \\[2mm]
Cox, P. H., Hellman,  W. S. and  Yildiz, A. (1984). Finite temperature corrections to field theory: electron mass, magnetic moment, and vacuum energy,  Ann. Phys. (N.Y.) {\bf 154}, 211.\\[2mm]
Dolan, L. and Jackiw, R. (1974). Symmetry behavior at finite temperature, 
Phys. Rev. D {\bf 9},  3320.\\[2mm]
Donoghue, J. F.,  Holstein, B. R. and Robinett, R. W. (1985). Quantum electrodynamics at finite temperature, Ann. Phys. (NY) {\bf 164}, No. 2, 233.\\[2mm]
Einstein, A. (1917). Zur Quantentheorie der Strahlung, Physikalische 
Zeitschrift {\bf 18}, 121.\\[2mm]
Haroche S. (2012). The secrets of my prizewinning research, Nature {\bf 490}, 311. \\[2mm]
Isihara, A.  Statistical mechanics, (Academic Press, New York,London, 1971).\\[2mm]
Johansson, A. E., Peressutti, G. and Skagerstam, B. S. (1986). Quantum field theory at finite temperature: renormalization and radiative corrections, 
Nucl. Phys. B {\bf 278}, 324.\\[2mm]
Pardy, M. (1989). Finite-temperature \v Cerenkov radiation, 
Phys. Lett. A {\bf 134},  No. 6,  357.\\[2mm]
Pardy, M. (1989). Finite-temperature gravitational \v Cerenkov radiation, 
International Journal of Theor. Physics {\bf 34}, No. 6, 951.\\[2mm]
Pardy, M. (1994). The two-body potential at finite temperature,
 CERN.TH.7397/94.\\[2mm]
Pardy, M. (2013a). Velocity of sound in the relic photon sea, arXiv: General Physics (physics.gen-ph)/1303.3201.\\[2mm]
Pardy, M. (2013b). Velocity of sound in the blackbody photon gas, 
Results in Physics {\bf 3}, 70.\\[2mm]
Partovi, H. M. (1994).  QED corrections to Planck’s radiation law and photon 
thermodynamics, Phys. Rev. D {\bf 50}, 1118.\\[2mm]
Planck, M. (1900). Zur Theorie des Gesetzes der Energieverteilung im Normalspektrum,
Verhandlungen deutsch phys. Ges. {\bf 2}, 237; ibid: (1901). Ann. Phys. {\bf 4}, 553.\\[2mm]
Rohlf, J. W. Modern physics from $\alpha \;to \; Z^{0}$, (John Wiley \& Sons LTD., London - New York,  1994).\\[2mm]
Sch\"opf, H-G. Theorie der W\"armestrahlung in historisch-kritischer Darstellung,
(Alademie/Verlag, Berlin, 1978).\\[2mm]
Sokolov, A. A., Loskutov, Yu. M. and Ternov, I. M. Quantum mechanics, (State Pedagogical Edition, Moscow, 1962). (in Russian). \\[2mm]
Weinberg, S. (1974). Gauge and global symmetries at high temperature, 
Phys. Rev. D {\bf 9}, 3357.\\[2mm]
Welton, Th. (1948). Some observable effects of the quantum-mechanical fluctuations of the electromagnetic field,  Phys. Rev. {\bf 74}, 1157.\\[2mm]

\end{document}